\documentclass[12pt,preprint]{aastex}

\def\simlt{\lower.5ex\hbox{$\; \buildrel < \over \sim \;$}}
\def\simgt{\lower.5ex\hbox{$\; \buildrel > \over \sim \;$}}

\def\cm3{{\rm cm^{-3}}}
\def\kms{km s$^{-1}$}
\def\msol{{M$_\odot$}}

\def\vexp{v_{\rm exp}}

\def\g11p2{G11.2$-$0.3}
\def\brg{Br$\gamma$\ }
\def\fetwo{[Fe II]\ }
\def\fetwoo{[Fe II]}
\def\mum{$\mu$m}
\def\tex{T_{\rm ex}}
\def\nh{n_{\rm H}}

\def\htwo{H$_2$\ }
\def\htwoo{H$_2$}
\def\fe164{[Fe II] 1.644~$\mu$m}
\def\hbeta{H$\beta$}
\def\h212{H$_2$ 2.122~$\mu$m}
\def\schi{{\sc Hi}\ }
\def\deltafe{\delta_{\rm Fe}}
\shortauthors{Koo et al.}
\shorttitle{\fetwo and \htwo filaments in G11.2$-$0.3}

\begin{document}

\title{\fetwo and \htwo filaments in the Supernova Remnant G11.2$-$0.3: Supernova Ejecta and 
Presupernova Circumstellar Wind}

\author{Bon-Chul Koo}
\affil{Astronomy Program, Department of Physics and Astronomy, Seoul National University, Seoul 151-747, Korea}\email{koo@astrohi.snu.ac.kr}

\author{Dae-Sik Moon\altaffilmark{1}}
\affil{Robert A. Millikan Fellow, Division of Physics, Mathematics, and Astronomy,
California Institute of Technology, MC 103-33, Pasadena, CA 91125}
\email{moon@srl.caltech.edu}

\author{Ho-Gyu Lee and Jae-Joon Lee}
\affil{Astronomy Program, Department of Physics and Astronomy, Seoul National University, Seoul 151-747, Korea}\email{hglee@astro.snu.ac.kr, jjlee@astro.snu.ac.kr}

\author{Keith Matthews}
\affil{California Institute of Technology, Downs Lab, MS 320-47, Pasadena, CA 91125}
\email{kym@caltech.edu}

\altaffiltext{1}{Department of Astronomy and Astrophysics, University of Toronto,
Toronto, ON M5S 3H4, Canada}

\begin{abstract}

We present the results of near-infrared imaging and spectroscopic 
observations of the young, core-collapse supernova remnant (SNR) 
G11.2--0.3. In the \fe164 image, we first discover 
long, clumpy \fetwo filaments within the radio shell of the SNR, 
together with some faint, knotty features in the interior of the remnant. 
The filaments are thick and roughly symmetric 
with respect to the NE-SW elongation axis of the central 
pulsar wind nebula. 
We have detected several \fetwo lines and \schi \brg line 
toward the peak position of the bright southeastern \fetwo filament. 
The derived extinction is large ($A_V=13$ mag) 
and it is the brightest \fe164 filament detected toward SNRs to date.
By analyzing two \fe164 images obtained in 2.2 yrs apart, we 
detect a proper motion corresponding to 
an expansion rate of $0.''035\pm 0.''013$ yr$^{-1}$ ($830\pm 310$~\kms). 
In addition to the \fetwo features, we also discover two
small \h212 filaments.
One is bright and along the SE boundary 
of the radio shell, while the other is faint 
and just {\em outside} of its NE boundary.
We have detected \htwo (2-1) S(3) line toward the former filament and 
derive an excitation temperature of 2,100~K.
We suggest that 
the \htwo filaments are dense clumps in
a presupernova circumstellar wind swept up by the SNR shock 
while the \fetwo filaments are probably composed 
of both shocked wind material and shocked supernova (SN) ejecta.
The distribution of \fetwo filaments 
may indicate that the SN explosion in \g11p2 
was asymmetric as in Cassiopeia A. 
Our results support the suggestion that \g11p2 is a remnant of 
a SN IIL/b interacting with 
a dense red supergiant wind. 

\end{abstract}


\section{Introduction}

G11.2$-$0.3 is a composite-type SNR with a central
pulsar wind nebula (PWN) surrounded by a circular shell.
The shell is bright both in radio and X-rays, and 
has an outer diameter of $4'$ and a thickness of $0.'5$ 
\citep{gre88,rob03}. 
The shell is clumpy with 
several clumps protruding its outer boundary.
The bright radio shell with high circular symmetry 
indicates that the remnant is young and it is thought 
to be the best candidate for the possible 
historical supernova of AD 386 \citep{ste02}.
The PWN with an associated pulsar was discovered  
at the very center of the remnant in X-rays with {\it ASCA}, 
and later its detailed structure 
was studied with the Chandra X-ray Observatory 
%\citep{vas96, tor97, tor99, kas01, rob03}. 
\citep[][and references therein]{vas96, kas01, rob03}. 
%The pulsar is located at the very center 
%of the remnant shell and its association with 
%the remnant appears definite.
%The pulsar, however, has a large characteristic age (24,000 yr) which 
%might indicate that the pulsar was
%only rotating as fast as it is now when it was born. (Roberts et al. 2003).
The PWN is elongated along the NE-SW direction with 
a total extent of $1'$, and appears to be surrounded by a radio 
synchrotron nebula with similar extent and shape.
The distance to G11.2$-$0.3 determined from \schi absorption 
is 5 kpc \citep{gre88}.
%distance HI absorption spectrum >26 kpc
%lack of strong absorption from +45 to the tangential point

The overall morphology of G11.2$-$0.3 resembles Cassiopeia A (Cas A). 
Both have a thick, bright, and clumpy shell, although 
the shell of Cas A is much brighter than that of \g11p2, i.e., 
2720 Jy vs 22 Jy at 1 GHz \citep{gre04}
\footnote{Also available at http://www.mrao.cam.ac.uk/surveys/snrs/.}.
At a distance of 3.4 kpc, the outer radius of Cas A shell is 2.0 pc
and its expansion velocity is 4,000--6,500 ~\kms\ \citep[e.g.,][]{fes96}, while 
they are 2.9 pc and $\sim 1,000$~\kms\ for G11.2$-$0.3. 
Cas A has a 
faint $5'$ (or 2.5 pc)-diameter plateau extending beyond the bright shell 
\citep[e.g.,][]{hwa04}. The plateau 
represents swept-up circumstellar (or ambient) medium while
the bright shell is thought to be mainly the ejecta swept up by 
a reverse shock.
Such plateau has not been detected in \g11p2. 
%Cas A has a central compact source object with no 
%observable pulsation and no surrounding nebula while 
%\g11p2 has a slowly-
\cite{che05} classified both 
Cas A and G11.2$-$0.3 into SN IIL/b category 
which has a red supergiant (RSG) progenitor star with some H envelope but
most lost \citep[cf.][]{hwa03, you06}. Detailed observations have  
revealed that the explosion of Cas A was 
turbulent and asymmetric and that the ejecta is now interacting  
with a clumpy circumstellar wind 
\citep[see][and references therein]{hwa04, che03}.
However, very little is known about the explosion and the interaction of 
the 1620(?) yr-old \g11p2 despite its close similarity to Cas A.

In this paper, we report the discovery and detailed studies of \fetwo and
\htwo filaments in the SNR \g11p2 using near-infrared (IR) imaging and spectroscopic 
observations. Although the recent mid-IR data obtained with the Spitzer Space 
Telescope show the 
presence of very faint wispy emission close to its SE boundary 
\citep{lee05, rea06}, our near-IR observations reveal much more prominent 
and extended features both at the boundary and interior of the remnant, which provide 
important clues on the origin and evolution of \g11p2.

\section{Observations}
 
We carried out near-IR imaging observations of the SNR G11.2-0.3 with 
Wide-field Infrared Camera (WIRC) on the Palomar 5-m Hale telescope 
using several narrow- and broad-band 
filters in 2003 June and 2005 August (Table 1). 
WIRC is equipped with a Rockwell Science Hawaii II 
HgCdTe 2K infrared focal plane array, covering $\sim 8.'5\times 8.'5$ field of view 
with $0.''25$ pixels scale. 
%Table 1 gives the summary of our imaging observations.
For the basic data reduction, we subtracted dark and sky background
from each individual dithered image and then normalized it by a flat image.
We finally combined the individual images to make a final image. 
The seeing was typically $0.''8$--$1''$ over the observations.
We obtained the flux calibration of our narrow-band filter (i.e., \fetwo and \htwo)
images using the H (for \fe164) and Ks (for \h212) band magnitudes
of $\ge 20$ nearby isolated, unsaturated
2MASS stars.  For this, we first converted the 2MASS magnitudes  
to fluxes \citep{coh03}, and then obtained the fluxes of the  
\fetwo and \htwo emission after we deconvolved the responsivities of both  
WIRC (\fetwo and \htwo) and 2MASS ($H$ and $K_s$) filters. The overall uncertainty in the  
flux calibration is less than $10 \%$.
We attribute the major source of uncertainty to the different  
band response of each filter as
the uncertainty in photometry itself is typically a few percent.  
For the astrometric solutions of our images, 
we used all the cataloged 2MASS stars in the field,  
and found that they are
consistent with that of 2MASS with rms uncertainty of $0.''15$.

After identifying several emission-line features in the 
aforementioned imaging observations,
we have carried out follow-up spectroscopic observations of them 
using Long-slit Near-IR 
Spectrograph of the Palomar 5-m Hale Telescope \citep{lar96} in 2005 August. 
The spectrograph has a $256\times 256$ pixel HgCdTe NICMOS 3 array 
with a fixed slit of $38''$ length. 
We placed the slit along the bright \fetwo and \htwo filaments crossing their peak positions. 
Toward the \fetwo filament, four spectra around 1.25, 1.52, 1.63 $\mu$m (for \fetwo emission), 
and 2.16 $\mu$m (for \schi \brg) were obtained, while, toward the \htwo filament, one spectrum around 
2.11 $\mu$m was obtained. Over the observations, the slit width was fixed to be $1''$, resulting in 
spectroscopic resolution of 650--850
with 0.06--0.12 $\mu$m usable wavelength coverage. For all the obtained 
spectra, the individual exposure time was 300 s with the same amount of exposure of nearby sky 
for sky background subtraction. For the \fetwo lines, we performed the exposure twice (with
different sky positions, but the same source position), and combined them, while, for the 
\htwo line, we performed the exposure only once. 

Just after the source observations, we obtained the spectra 
of the G3V star HR 8545, which was at the similar airmass of the source 
by uniformly illuminating 
the slit using the f/70 chopping secondary of the telescope. We then divided the source spectra 
by those of HR 8545 and multiplied by a blackbody radiation curve of the G3V star temperature, 
which is equivalent to simultaneous flat fielding and atmospheric opacity correction. 
G stars, however, have numerous intrinsic (absorption) feautres, 
so that this procedure could inflate the intensities of emission lines 
if they fall on these stellar features.
%The correction is not straightforward because the strength of intrinsic features vary 
%considerably with spectral type and stellar parameters 
% \citep{mai96;vac03}. 
We have estimated the errors using 
the G2V solar spectrum \citep{liv91}\footnote{Also
available at  http://diglib.nso.edu/contents.html.}
as a template of our reference star HR 8545 \citep[cf.][]{mai96,vac03}. 
The estimated errors in the observed line fluxes are 
$\le 5$\% for all lines except \brg line for which it is $10$\%.
The resulting errors in the line ratios, which are used for the 
derivation of physical parameters, are 
$\le 2$\% except for the \brg/\fe164 ratio for which it is 
7\%.  These calibration errors 
are all less than their statistical ($1\sigma$) errors (see Table 2). 
Another source of error is different atmospheric condition. 
All spectra of the source and HR 8545 were obtained at air mass 
of 1.6--1.8 except the H$_2$ spectra of the source which was 
obtained at an air mass of 2.27.
There is a strong atmospheric CO$_2$ absorption line between 2.05 and 2.08 $\mu$m, and  
the different airmasses can give an error 
in the intensity of H$_2$ (2--1) S(3) line at 2.0735 $\mu$m.
According to \cite{han96}, the error is about $2$\%  
when the air masses differ by 0.3, so that it would be 
$\simlt 5$\% for our H$_2$ (2--1) S(3) line. Again, this is less
than the ($1\sigma$) statistical error. 
We therefore consider that the uncertainty due to the calibration errors
is less than the statistical errors quoted in this paper (Table 2). 
For the
wavelength solutions of the spectra, we used the OH sky lines \citep{rou00}.

\section {Results}

Fig. 1 (right) is our three-color image representing the near-IR
\fe164 (B), \h212 (G), and \brg 2.166 $\mu$m (R)  emission
of the SNR \g11p2. We also show 
an 1.4 GHz VLA image for comparison which 
was obtained by \cite{gre88} in 1984--85 with 
$3''$ resolution.
%\footnote{There is a more recent observation made by 
%\cite{tam03} during 2001--2002, but ...
Note that the expansion rate of \g11p2 at 1.4 GHz is 
$0.''057\pm 0.''012$ yr$^{-1}$ \citep{tam03} which amounts to 
$\sim 1''$ over the last 20 years (see also \S 3.3). 
The near-IR emission features in Fig. 1 can be summarized as follows:
(1) an extended ($\sim 2.'5$), bright \fetwo (blue) filament along the 
SE radio shell;  
%\footnote{This is in fact the brightest among the known \fe164
%filaments associated with SNRs (see 3.1).};
(2) some faint, knotty \fetwo emission features along the NW radio shell
as well as in the interior of the source;
(3) a small ($30''$), bright \htwo (green) filament along the outer boundary
of the source in the SE;
(4) another small, faint \htwo filament outside the NE bounday of
the source.
Overall the \fetwo filaments are located either within the radio shell
or inside of the source, while the \htwo filaments are along the radio boundary
or even outside of it. We have not found any apparent \brg filament in our
rather shallow imaging observation, although we have detected faint 
\brg line
emission toward the \fetwo peak position in our spectroscopic observation.
In the following, we summarize the results on the \fetwo and 
\htwo emission features.

\subsection{\fe164 emission}
\subsubsection{Photometry}

In order to see the \fetwo emission features more clearly, we 
have produced an `star-subtracted' image (Fig. 2).
%We used H-cont image to subtract stars. 
%We first aligned the H-cont and \fe164 images using bright stars 
%within 
We first performed PSF photometry of H-cont and \fe164 images, and 
removed stars in the \fe164 image if they had
corresponding ones in the H-cont image.
This PSF photometric subtraction left residuals around 
bright stars which we masked out. 
The faint stars, which were not removed by the PSF subtraction because
the H-cont image is not as deep as that of \fetwo, were then removed
by subtraction of median value of $15 \times 15$ nearby pixels.
Fig. 2 is the final star-subtracted image where we can see the detailed 
features of \fetwo emission more clearly.

As in Fig. 1 (right), the extended filament within the southeastern SNR shell, 
\fetwoo-SE filament hereafter, is most prominent. 
The filament is composed of two 
bright, $30''$-long, elongated segments in the middle and 
two clumpy segments at the ends. The one at the southern end is 
a little bit apart from the other three. 
The total extent of the filament is $\sim 2.'5$. 
The filament is not very thin but has a width of $\simlt 10''$.
Fig. 3 shows a detailed structure of the filament, where 
we have just masked out stars using the K-cont image 
in order to avoid any possible artifacts associated with 
the PSF photometric subtraction.
We can see that the filament has a very good correlation with the radio shell both 
in morphology and brightness. 
The peak \fe164 surface brightness of the filament is 
$1.9 \pm 0.2\times 10^{-3}$ ergs cm$^{-2}$ s$^{-1}$ sr$^{-1}$, which 
is larger than any previously reported 
brightness of \fe164 filaments in other remnants, e.g., 
1.1--3$\times 10^{-4}$~ergs cm$^{-2}$ s$^{-1}$ sr$^{-1}$ in 
IC 443 and Crab \citep{gra87, gra90} or 
$1.5\times 10^{-3}$~ergs cm$^{-2}$ s$^{-1}$ sr$^{-1}$ in 
RCW103 \citep{oli89}. 

On the opposite side of the SE filament lies another long ($\sim 2.'5$) 
filament within the northwestern SNR shell (Fig. 4).
This filament (\fetwoo-NW filament) is relatively faint 
and appears to be clumpy. 
%The bright clump of $3.''5$-extent 
%at (18 11 24.11, -19 24 13.2) is noticeable.
It has little correlation with the radio emission.
%, e.g., the eastern half traces the region with enhanced radio emission.
We note that \fetwoo-SE and NW filaments lie roughly 
symmetric with respect to the line of position angle 
$\approx 60^\circ$, which is close to the inclination 
%($\approx 64^\circ$) 
of the central PWN of \g11p2 in X-ray \citep{rob03}.
In addition to these two extended filaments, some faint, knotty 
emission features are also seen 
in the interior of the remnant, particularly in the southern area (Fig. 5).
These features spread over an area of 
$\sim 2'$ extent and filametary,
with some of them having a partial ring-like structure. There are
also several bright clumps of $\sim 5''$ size. 
Most of the clumps appear to
be connected to the filaments, although some are rather isolated.
The brightnesses of these central emission features and NW filament 
are $\simlt 3\times 10^{-4}$~ergs cm$^{-2}$ s$^{-1}$ sr$^{-1}$.
The observed total \fe164 flux is estimated to be 
$1.1\pm 0.2 \times 10^{-11}$~erg cm$^{-2}$ s$^{-1}$,
$76\pm 12$\% of which is from the SE filament.
%The observed peak \fe164 surfcace brightness of the SE filament is 
%$1.94 \pm 0.19\times 10^{-3}$ ergs cm$^{-2}$ s$^{-1}$ sr$^{-1}$.
%This is larger than the brightness of \fe164 filaments in other remnants, e.g., 
%IC 443 ($1.1\times 10^{-4}$~ergs cm$^{-2}$ s$^{-1}$ sr$^{-1}$, \cite{gra87}), 
%RCW103 ($1.5\times 10^{-3}$~ergs cm$^{-2}$ s$^{-1}$ sr$^{-1}$, \cite{oli89}),
%Crab ($3\times 10^{-4}$~ergs cm$^{-2}$ s$^{-1}$ sr$^{-1}$, \cite{gra90}). 
%The photometric properties of the filaments are summarized in Table 2.

%Fig.3 shows the one-dimensional intensity distribution along 
%the line crossing the peak position of the \fetwoo-SE filament. 
%The peak is located right in the middle of the radio 
%filament. The peak surfcace brightness is 
%$1.47\times 10^{-3}$ erg cm$^{-2}$ s$^{-1}$. 
%implies Fe II column density of 
%$\nhtwo=5\times 10^{20}$~cm$^{-2}$.

\subsubsection{Spectroscopy}

We have detected several \fetwo lines toward the peak position of 
the \fetwoo-SE filament (\fetwoo-pk1). Table 2 summarizes the 
detected lines and their relative strengths, and Fig. 6 shows the 
spectra. \fetwo 1.257~$\mu$m and \fe164 lines originate from the 
same upper level, so that their unreddened flux ratio is fixed by 
relative Einstein $A$ coefficients which is 1.04 according to 
\cite{qui96}. Toward \fetwoo-pk1, the ratio is 0.31, which implies 
$A_V=13$ mag  ($A_{1.644\mu {\rm m}}=2.43$ mag) or H-nuclei 
column density of $2.49\pm 0.07 \times 10^{22}$~cm$^{-2}$ using 
the extinction cross section of the carbonaceous-silicate model for 
interstellar dust with $R_V=3.1$ of \cite{dra03}\footnote{Data 
available at http://www.astro.princeton.edu/~draine/dust/dustmix.html.}.
This is a little larger than the
column density to the remnant derived from
X-ray observations $(1.7-2.4)\times 10^{22}$~cm$^{-2}$ \citep{rob03}.
We note that the numerical values of the Einstein $A$ coefficients 
for near-IR \fetwo lines in the literature differ as much as 50\%: 
using the values of \cite{nus88}, the expected \fetwo 1.257~$\mu$m to 
\fe164 line-intensity ratio is 1.36, while \cite{smi06} empirically 
derived 1.49 from their spectroscopy of P Cyg. If the intrinsic ratio 
is 1.36 or 1.49, we obtain a little (20--30\%) higher column density. 
We adopt the $A$-values of \cite{qui96} in this paper which yield a 
column density closer to the X-ray one. 
According to \cite{har04}, they also yield extinction more consistent 
with optical spectroscopic result for a protostellar jet.
%The derived extinction toward \fetwoo-pk1 yields 
%the dereddened \fe164 peak surface brightness of 
%$1.82 \pm 0.18\times 10^{-2}$ ergs cm$^{-2}$ s$^{-1}$ sr$^{-1}$. 
%The total dereddened \fe164 flux is 
%$1.0\pm 0.1 \times 10^{-10}$~erg cm$^{-2}$ s$^{-1}$.

The ratios of the other three lines, e.g., \fetwo 1.534~\mum,
1.600~\mum, and 1.664~\mum, to \fetwo 1.644~\mum\
are good indicators of electron density \citep[e.g.,][]{oli90}.
%Their dereddened ratios are $0.152\pm 0.005$, $0.113\pm 0.003$, and $0.050\pm0.002$, 
%respectively.  
We solved the rate equation using the atomic parameters 
assembled by CLOUDY \citep[version C05.05,][]{fer98} which adopts 
the Einstein A coefficients of \cite{qui96} and 
collision strengths of \cite{pad93} and \cite{zha95}. We have included 
16 levels which is enough at temperatures of our interest ($\simlt 10^4$~K).
We consider only the collisions with electrons, neglecting those 
with atomic hydrogen, even if the degree of 
ionization of the emitting region could be low (see \S~4.2). 
This should be acceptable since the rate coefficients for atomic hydrogen collisions
are more than two orders of magnitude smaller than those for electron 
collisions \citep{hol89}.
The ratios of 1.534~\mum\ and  1.664~\mum\ lines
yield consistent results, e.g., $6,000\pm 400$~cm$^{-3}$ and
$5,900\pm 400$~cm$^{-3}$, while 1.600~\mum\ line ratio yields a little
higher density ($7,800\pm 400$~cm$^{-3}$) at $T=5,000$~K which is 
the mean temperature estimated for \fetwo line-emitting regions in other SNRs
\citep[][; see also \S~4.2]{gra87, oli89}. The result is
not sensitive to temperature, e.g., a factor of 2 variation in temperature
causes 10--20\% in density.
We adopt the average value $6,600\pm 900$~cm$^{-3}$ at $T=5,000$~K  
as the characteristic electron density of the \fetwo filaments.

We also detected \brg line toward \fetwoo-pk1. 
The dereddened ratio of \fe164 to \brg line is $77^{+14}_{-10}$, 
which is much greater than that ($\simlt 0.1$)
of HII regions but comparable to the ratios observed in other SNRs
(see \S~4.2).

\subsubsection{Proper Motion during 2003--2005}

We have two \fe164 images taken in 2.2 years apart, i.e., in 
2003 June and 2005 August. The time 
interval is not long enough to notice the proper motion of 
the \fetwo filaments in the difference image obtained by 
subtracting one from the other.
We instead inspect 
one-dimensional intensity profiles of the 
bright \fetwo-SE filament to search for its proper motion 
associated with an expansion.

Fig. 7 shows the intensity profiles across the two bright
segments of the \fetwoo-SE filament along the cuts (dashed lines) in Fig. 3. 
The cuts are made to point to the central pulsar which is 
very close to the geometrical center of the SNR shell \citep{kas01}. 
The distance in the abscissa is measured from the upper right end of the cuts, 
so that it increases outward from the remnant center. 
Note that the
profiles of the filament in 2005 (solid lines) are
slightly shifted outward from those in 2003 (dashed lines). 
We fit the profiles along the cuts A and B with a Gaussian and obtain
shifts of $0.''063\pm 0.''032$ and $0.''095\pm 0.''064$ in 
their central positions, respectively. 
For comparison, the profiles of nearby 
stars, e.g., the strong peak at $24''$ in Fig. 7 (left),
do not show any appreciable shift.
The mean shift in stellar positions from the same 1-dimensional Gaussian 
analysis of nearby seven stars is found to be $-0.''0067 \pm 0.''0029$. 
Therefore, the mean proper motion 
of the SE filament with respect to the 
nearby stars during 2.2 years amounts to $0.''076 \pm 0.''029$, which 
corresponds to a rate of $0.''035\pm 0.''013$ yr$^{-1}$.
%at a distance of 5 kpc, an expansion velocity of $830\pm 310$~\kms.
%For comparison \cite{tam03} obtained an expansion rate of 
%$0.''057\pm 0.''012$ yr$^{-1}$ at 1.465 GHz and 
%$0.''040\pm 0.''013$ yr$^{-1}$ at 4.860 GHz 
%by comparing radio images separated by 17 years.

\subsection{\h212 emission: Photometry and Spectroscopy}

Fig. 8 is a star-subtracted and median-filtered \h212 image. 
The image has been made in the same way as Fig. 2. 
Two small ($\sim 30''$) filaments, one at the 
southern SNR radio boundary and another fainter one outside the NE 
boundary are now clearly seen.
The one in the southeast (\htwo-SE filament) is bright 
and elongated along the radio boundary. 
Its peak surface brightness is 
$3.0\pm 0.3  \times 10^{-4}$ erg cm$^{-2}$ s$^{-1}$ sr$^{-1}$ 
and its flux is $4.3 \pm 0.4 \times 10^{-13}$~erg cm$^{-2}$ s$^{-1}$.
The NE filament (H$_2$-NE filament) is just outside of the SNR boundary 
and is located where the radio continuum boundary is distorted.
Its
surface brightness is $\simlt 40$\% of the SE filament peak brightness,
while its flux density is $\sim 50$\% of the SE filament. 
There is no \fe164 emission associated with either \htwo filament. 
A long ($\sim 2'$) filamentary feature seems to be present well outside the 
southeastern SNR boundary, but it is too faint to be confirmed.

Fig. 9 shows a detailed structure of the \htwoo-SE filament.
It is composed of two bright segments surrounded by a diffuse envelope.
It is just outside of the bright \fetwoo-SE filament, but there is no 
apparent correlation between the two (cf. Fig. 3).
We have detected two \htwo lines, (1,0) S(1) and (2,1) S(3), 
toward the peak position of 
the filament, \htwoo-pk1 (Fig. 10). Their dereddened ratio, using the 
column density derived from \fetwo line ratios 
($A_{2.12 \mu {\rm m}}=1.59$ mag), is 
$0.14\pm 0.01$ (Table 2), which gives $\tex\approx 2,100$ K 
using the transition probabilities of \cite{wol98}.
%The dereddened peak \h212 surface brightness is 
%$1.3\pm 0.1 \times 10^{-3}$ erg cm$^{-2}$ s$^{-1}$ sr$^{-1}$, and the 
%dereddened total flux of the SE filament is 
%$1.9\pm 0.2 \times 10^{-12}$ erg cm$^{-2}$ s$^{-1}$. 

\section{Discussion}

G11.2$-$0.3 has been proposed to be 
a young remnant of an SN IIL/b interacting with a dense RSG wind 
based on its PWN and the small size of the SNR shell \citep{che05}. 
The thick, bright shell is thought to be shocked SN ejecta 
in contact with shocked wind material. 
The outer edge of the shell is not sharp and 
it was suggested that 
the ambient shock propagating into wind material 
could be at a larger distance \citep{gre88,che05}.
In the following, we first discuss the 
physical properties of the \htwo filaments that we have discovered 
in this paper, and show that our results support the SN IIL/b scenario. 
Then we discuss the physical properties of the \fetwo filaments
which are thought to be composed of both shocked wind 
material and shocked SN ejecta.

\subsection{\htwo Filaments and Presupernova Circumstellar Wind}
\subsubsection{Excitation of \htwo filaments}

The \htwoo-SE filament is located at the rim of the bright SNR shell and 
elongated along the rim, which 
suggests that it is excited by the SNR shock.
The derived $v=2$--1 excitation temperature ($\approx 2,100$~K)
is also typical for shocked molecular gas \citep{bur89}. 
The dereddened peak \h212 surface brightness is 
$1.3\pm 0.1 \times 10^{-3}$ erg cm$^{-2}$ s$^{-1}$ sr$^{-1}$, and the 
dereddened total flux of the SE filament is 
$1.9\pm 0.2 \times 10^{-12}$ erg cm$^{-2}$ s$^{-1}$. 

The interstellar ultraviolet (UV) photons in 
principle could excite and heat the \htwo gas to 
produce similar excitation temperature if the gas is dense enough 
for collisions to dominate deexcitation \citep{ste89, bur90}.
However, the expected \h212 surface brightness by UV photon excitation 
is low unless 
the density is high and the radiation field is very strong, 
e.g., $\nh\ge 10^5$~cm$^{-3}$ and $G_0\ge 10^4$ for 
$\ge 1\times 10^{-4}$~erg cm$^{-2}$ s$^{-1}$ sr$^{-1}$ where 
$\nh$ is the number density of H nuclei and 
$G_0$ is far UV (FUV) intensity relative to the interstellar 
radiation field in the solar neighborhood \citep{bur90}. 
Note that $G_0=10^4$ corresponds to an O4-type star at a distance of 
$\sim 1$ pc \citep{tie05}.
No such strong FUV source exists around the filament.

X-ray emission from the remnant is another source 
that could possibly excite and heat the \htwo filament. 
We may consider a 
molecular clump situated at some distance from an SN explosion.
As the SN explodes and the SNR evolves, 
the X-ray flux increases and, in principle, 
an ionization-dissociation front may develop and propagate into 
the clump. If the density is sufficiently high, the \htwo lines from  
heated molecular gas could show `thermal' line ratios \citep{gre95}.
The \htwo line intensities from such clump 
depend on details, and no model calculations that may be directly 
applicable to 
our case are available \citep[cf.][]{dra90, dra91, mal96}.
In the following, we instead simply consider the energy budget.
If the \h212 line is emitted by reprocessing the X-ray photons 
from the SNR falling onto the molecular clump, its luminosity may be written as 
$L_{2.122} \sim \epsilon L_X (\Omega_{\rm cl}/4\pi)$ where 
$\epsilon$ is an efficiency of converting the incident 
X-ray energy flux into 
\h212 line emission, $L_X$ is the X-ray luminosity of the remnant,
and $\Omega_{\rm cl}$ is the solid angle of the 
clump seen from the SNR center. The above formula is accurate if 
the clump is small and if the X-ray source is 
spherically symmetric. Although \g11p2 is 
not a spherically symmetric source in X-rays, we may use the formula to 
make a rough estimate of the expected  \h212 line luminosity.
The conversion efficiency 
for SNRs embedded in molecular clouds  
was calculated to be 
$\simlt 1 \times 10^{-3}$ \citep{lep83, dra90, dra91}. 
The efficiency is a function of X-ray energy absorbed 
per H-nucleon and the above inequality might be valid 
for X-ray irradiated small clumps too. 
Now if we assume that the \htwo clump has the line-of-sight extent 
similar to the extent on the sky ($\sim 0.'5$), then
$ \Omega_{\rm cl}/4\pi \sim 4 \times 10^{-3}$.
Since the X-ray luminosity of \g11p2  
is $L_X\sim 10^{36}$~erg s$^{-1}$ in 0.6--10 keV band \citep{vas96}, we have 
$L_{2.122}\simlt 4 \times 10^{30} $~erg s$^{-1}$. 
This is much 
less than the observed \h212 
luminosity of the SE filament, 
which is $\sim 6 \times 10^{33}$~erg s$^{-1}$.
Therefore, the X-ray excitation/heating 
does not appear to be important for the \htwoo-SE filament.  

The above consideration leads us to conclude that 
the \htwoo-SE filament is excited by the SNR shock
associated with \g11p2.
The absence of associated 
\fe164 or \brg emission suggests that the \htwo emission 
from the \htwoo-SE filament might be 
from warm molecules swept-up by 
a slow, non-dissociative $C$ shock not from reformed molecules
behind a fast, dissociative $J$ shock. 
%The column density required to 
%reform \htwo molecules is also large,  \citep{hol89}.
The critical velocity for a shock to be a non-dissociative 
$C$ shock is $\simlt 50$~\kms\ \citep{dra83, mck84}. 
The dereddened mean surface brightness of the \htwoo-SE filament is   
$\sim 8\times 10^{-4}$~erg cm$^{-2}$ s$^{-1}$ sr$^{-1}$. 
This is comparable to the (normal) brightness of 
a $\sim 30$~\kms\ shock propagating into molecular gas of 
$n_{\rm H}=10^4$~cm$^{-3}$ according to the 
$C$-shock model of \cite{dra83}. 
We were unable to find model calculations for lower densities.
But, since the intensity will be proportional 
to the preshock density, provided that the density 
in the emitting gas is less than the critical density 
\citep[$\simgt 10^5$~cm$^{-3}$;][]{bur89}, 
%Burton et al. 1989, MNRAS, 236, 409
%Richter et al. 1995, ApJ, 454, 277
the results of \cite{dra83} indicates that 
a 40--50~\kms\ shock propagating into 
molecular gas of $n_{\rm H}=10^3$~cm$^{-3}$ might have 
similar (normal) surface brightness. 
A slower shock with a
lower preshock density would be
possible if the shock propagating into the \htwo filament
is tangential along the line of sight, so that
the brightness normal to the shock front is lower.
 
%The derived excitation 
%temperature is consistent with these shock parameters too.
%although we need 
%H$_2$ lines over broad energy levels for shock diagnostics.

The situation is not so clear 
for the \htwoo-NE filament for which we lack spectroscopic information.
Its flux density, however, is comparable to that of the SE filament and 
we may rule out the excitation by X-rays from \g11p2. 
We checked 2MASS colors of nearby ($\le 2'$) stars, but found 
no OB stars that would be responsible for the UV excitation.
This leaves again the shock excitation for the origin of the 
\htwo emission.
A difficulty with the shock excitation is that 
the filament is located outside the radio SNR boundary. 
But as have been pointed out in previous studies \citep[e.g.,][]{gre88},
the radio continuum boundary is not sharp and 
the ambient shock is thought to have propagated beyond the apparent radio 
boundary. It therefore seems to be reasonable to consider that 
the \htwoo-NE filament is excited by the SNR shock too, although
we need spectroscopic observations to understand the 
nature of the \htwoo-NE filament.

\subsubsection{Circumstellar Origin of \htwo filaments}

The \htwo filaments are more likely of circumstellar origin than interstellar. 
If interstellar, 
they must be dense clumps originally in an ambient or parental molecular cloud. 
We do not expect to observe molecular material 
around small, young core-collapse SNe in general because massive stars 
clear out the surrounding medium with their strong UV radiation and 
strong stellar winds during their lifetime. Some molecular material may 
survive if the progenitor star is an early B-type (B1--B3) star, which does not have 
strong UV radiation nor strong stellar winds 
\citep{mck84b, che99}. A difficulty with this scenario, however, is that 
then the swept up mass at the current radius (3 pc) 
is likely to be much greater than the 
ejecta mass, so that the remnant should have been already in Sedov 
stage where it would appear as a thin, limb-brightened shell. 
The thick-shell morphology of \g11p2, however,  
indicates that it is not yet in Sedov stage. 
We therefore consider that the \htwo filaments 
are of circumstellar origin which fits well into the SN IIL/b scenario. 

It is plausible that the progenitor of \g11p2 had a strong wind 
which contains dense clumps.
Numerous such clumps have been observed in Cas A, e.g., 
``Quasi-stationary flocculi (QSF)", which are 
slowly moving, dense optical clumps immersed within 
a smoother wind \citep{van71, van85}. In the 320-yr old Cas A, 
the shock propagating into the clump is 
fast \citep[100--200~\kms;][]{che03} while in the 1620-yr old \g11p2 it is slow 
(30--50~\kms). Their velocity ratio is comparable to 
the ratio ($\sim 1/5$) 
of SNR expansion velocities, which suggests 
that the winds in \g11p2 and Cas A have similar properties.
We may estimate the density contrast 
between the clump and the smoother wind from 
the ratio of the shock speed into the clump ($v_c=30-50$~\kms) to 
the SNR forward shock speed $\vexp$. If we adopt the result of 
the radio (20 cm) expansion studies by \cite{tam03},  
$\vexp=1350\pm280$~\kms\ so that the density contrast would be 
$(\vexp/v_c)^2=700-3,000$. 
For comparison, \cite{che03} estimated a density contrast of $3,000$ for Cas A.
%This might be a lower limit
%because the forward shock speed could be greater than
%the current expansion velocity of the shell.
%These shocked clumps will be swept up by 
%the bright SNR shell, which might explain why we observe only 1--2 
%shocked \htwo clumps 

\subsection {\fetwo Filaments and SN Ejecta}
\subsubsection {Shock Parameters of the \fetwoo-SE filament}

The \fetwo filaments are located within the bright SNR shell 
in contrast to the \htwo filaments. 
The \fetwoo-SE filament  
has a remarkable correlation with the radio shell in both morphology and 
brightness.
The knotty emission features inside the remnant might be within the 
shell too, but projected on the sky. 
The location of the filaments 
suggests that the \fetwo emission is almost certainly from the shocked gas. 
The shock must be radiative and 
the \fetwo emission should originate from the cooling layer
behind the shock.

The \fetwoo-SE filament is very bright 
with the dereddened \fe164 peak surface brightness of 
$1.80 \pm 0.18\times 10^{-2}$ ergs cm$^{-2}$ s$^{-1}$ sr$^{-1}$. 
It is in fact the brightest among the known \fe164
filaments associated with SNRs.
The total dereddened \fe164 flux is 
$1.0\pm 0.1 \times 10^{-10}$~erg cm$^{-2}$ s$^{-1}$.
%We may derive the shock parameters of the \fetwoo-SE filament 
%from the observed properties. 
The ratio of \fe164 to \brg line ($\sim 80$) toward 
the peak position of the \fetwoo-SE filament 
is larger or comparable to the ratios observed in other SNRs, 
e.g., 27 to $\ge 71$ in IC 443 \citep{gra87} or 34 in RCW 103 \citep{oli89}.
It was pointed out in previous studies that the 
high ratio can result from SNR shocks {\em interacting with the ISM}  
by the combined effects of 
`shock excitation' and the enhanced gas-phase iron abundance. 
First, since the ionization potential of iron atom is only 7.9 eV, 
FUV photons from the hot shocked gas 
can penetrate far downstream to 
maintain the ionization state of Fe$^+$ 
where H atoms are primarily neutral
\citep{mck84, hol89b, oli89}. Therefore, \fetwo lines 
are emitted mainly in gas with 
a low degree of ionization at $T=10^3$--$10^4$~K.
This partly explains the observed high ratio of \fe164 to \brg lines,
but not all. Shock model calculations showed that 
the ratio is $\sim 1$ if the gas-phase iron abundance 
is depleted as in normal ISM.
According to \cite{hol89b}, the ratio is $\sim 1.5$ for 
shocks at velocities 80--150~\kms\ propagating into 
a molecular gas of $\nh=10^3$~cm$^{-3}$ with iron depletion 
$\deltafe\equiv{\rm [Fe/H]/[Fe/H]_\odot}=0.03$ where 
[Fe/H]$_\odot$=$3.5\times 10^{-5}$.
\cite{mck84} presented the results on atomic shock calculations 
including grain destruction: for a 100~\kms\ shock
propagating into atomic gas of $\nh=10$ and 100~cm$^{-3}$, 
[Fe II] 1.2567 $\mu$m/\hbeta=2.7 and 3.7 with 
$\deltafe=0.53-0.58$ in the far downstream. 
If we use 0.033 as the ratio of 
%1.04 as the ratio of \fe126 to \fe164 line intensities and 
\brg to \hbeta\ line intensities which corresponds to 
a Case B nebula at 5,000 K \citep{ost89}, 
the ratio corresponds to \fe164/\brg=80 and 110, comparable to the observed 
ratio. Therefore, gas-phase iron abundance close to 
the solar is required 
to explain the observed \fe164 to \brg ratio toward \fetwoo-pk1.

%edge-on
The preshock density may be estimated from the 
\fe164 brightness.
The \fe164 surface brightness toward the \fetwoo-SE filament varies 
$\sim 1-10\times 10^{-3}$ erg cm$^{-2}$ s$^{-1}$ sr$^{-1}$.
Its morphology in Fig. 4 suggests that 
the shock front might be tangential along the line of sight to 
enhance the surface brightness of the filament.
The normal surface brightness  
of the 100~\kms\ shock
propagating into atomic gas of $\nh=100$~cm$^{-3}$ is 
$2.5\times 10^{-4}$ erg cm$^{-2}$ s$^{-1}$ sr$^{-1}$
\citep{mck84}. 
It is $0.3-2\times 10^{-3}$ erg cm$^{-2}$ s$^{-1}$ sr$^{-1}$
for 80--150~\kms\ shocks propagating into 
a molecular gas of $\nh=10^3$~cm$^{-3}$
if the gas-phase abundance of iron was solar \citep{hol89b}.
Therefore, 
%even considering the projection effect, 
the preshock density needs to be $\simgt 1,000$ cm$^{-3}$.
This appears to be roughly consistent with the electron
density derived from \fetwo lines ratios. 
As we pointed out above, the ionization fraction of the 
\fetwoo-emitting region is expected to be low.
\cite{oli89} estimated a mean ionization fraction of 0.11, in
which case $\nh\approx n_e/0.11\approx 6\times 10^4$~cm$^{-3}$. 
For a 100~\kms\ shock, the final compression factor would be 
$\sim 80$ \citep{hol89}, so that the above postshock 
density implies a preshock density of $\sim 800$~cm$^{-3}$. 
This is close to the density required to explain the surface 
brightness considering the uncertainties in various parameters. 
Therefore, a 100~\kms\ shock propagating into a gas 
of $\nh\simgt 1,000$ cm$^{-3}$ and destroying dust grains 
seems to explain the observed parameters of the \fetwoo-SE filament. 

\subsubsection{Origin of \fetwo filaments}

The \fetwo filaments could be either shocked circumstellar medium (CSM) 
or shocked ejecta, or both under the context of the Type IIL/b scenario. 
In the SE filament, 
\schi \brg line is detected at the peak position and 
its ratio to \fe164 line is consistent with a $100$~\kms\ 
{\em interstellar} shock (\S~4.2.1), which implies that 
the emission is not from metal-rich ejecta but from shocked CSM.  
For example, when the \htwo clumps 
in the previous section are swept up by shocked dense ejecta, a stronger
shock will propagate into the clumps to dissociate and ionize 
the gas to produce \fetwo emission. 
Radio observation also 
suggests that the remnant is more heavily affected by the 
ambient medium toward this direction: \cite{kot01} showed that 
the magnetic field structure of the bright radio shell is radial 
in general except the bright SE shell where the 
degree of polarization is significantly low compared to the other parts
of the shell. The non-radial magnetic field and the low 
degree of polarization suggest that the synchrotron emission 
is dominated by shocked ambient gas not by shocked ejecta. 
On the other hand, the SE filament is located in the 
middle of the radio shell and has a large radial proper motion.
If the proper motion is due to expansion of the SNR shell,
which is very likely, it implies 
an expansion velocity of $\ge 830\pm 310$~\kms\ (see next).
This suggests that the filament is associated with ejecta. 
It is possible that some \fetwo emission originates from dense, Fe-rich ejecta 
recently swept-up and excited by a reverse shock.
We suppose that the \fetwoo-SE filament consists of 
both the shocked CSM and the shocked ejecta, although it is not 
obvious how the two interact to develop the observed properties. 

The derived proper motion of the \fetwoo-SE filament 
($0.''035\pm 0.''013$ yr$^{-1}$) may be compared to the expansion rate 
of the radio shell. \cite{tam03} obtained a mean expansion rate of 
$0.''057\pm 0.''012$ yr$^{-1}$ at 1.465 GHz and 
$0.''040\pm 0.''013$ yr$^{-1}$ at 4.860 GHz 
by comparing radio images separated by 17 years.
%They derived the expansion rates of individual conic sections 
%of the shell, and the expansion rate of the 
%SE shell was not noticeably different from the overall expansion rate.
Our proper motion is comparable to the 4.860-GHz expansion rate 
but is smaller than the 1.465-GHz expansion rate which 
was considered to be more reliable by the authors. It is possible that 
the \fetwoo-SE filament is not moving perpendicularly to the sight line, 
so that the true space motion is greater. But, 
considering that the filament is located near the boundary of 
the remnant, the projection effect is probably not large.
Instead the difference may be because 
the proper motion that we have derived in this paper represents the 
velocity of the brightest portion of the filament while 
the radio expansion rate might be close to the pattern speed, 
e.g., the SNR shock speed. 
Since the velocities of shocked ambient gas and shocked ejecta 
in the shell might be less than the SNR shock velocity,
it is plausible that our `expansion rate' is less than the radio one.
We will explore the dynamical properties of 
\g11p2 in our forthcoming paper. 

%, e.g., $0.9\pm0.5$\% vs 5--10\%. \cite{kot01} postulated that 
The \fetwoo-NW filament and the knotty emission features 
are considered to be mostly, if not all, dense SN ejecta.
The radial magnetic field 
supports this interpretation \citep{kot01}.
Their filamentary and ring-like structure 
may be a consequence of bubbly Fe ejecta \citep[e.g.,][]{blo01}.
It is worth to note that the \fetwo emission is distributed mainly 
along the NW-SE direction (Fig. 2), the direction perpendicular 
to the PWN axis. 
The long and symmetric morphology of the \fetwoo-SE and -NW filaments 
resembles the main optical shell of Cas A.
Cas A, in optical forbidden lines of O, S ions,
shows a complex northern shell composed of several
bright, clumpy filamentary structures at varying distances from the center and
a relatively simple-structured southern shell \citep[e.g.,][]{fes01}.
These northern and southern portions of the main optical shell
are opposite across the jet-axis along the NE-SW axis.
The optical shell is generally believed to be
dense clumps in ejecta recently swept up by reverse shock, although 
it contains QSFs too.
The similarity to Cas A suggests that the explosion in \g11p2 was 
asymmetric as in Cas A. 

The total \fe164 luminosity is 
$\sim 75 L_\odot$. This is two orders of magnitude greater than 
Kepler or Crab, but comparable to RCW 103 or IC 443 
\citep{oli89, kel95}. 
In collisional equilibrium at $T=5,000$~K 
with $n_e\approx 6,600$~cm$^{-3}$, this converts to Fe mass of 
$\sim 5.3\times 10^{-4}$~\msol. Both the shocked ejecta and the 
shocked CSM constitue this. The $^{56}$Fe mass 
that would have been formed from the radioactive decay of 
$^{56}$Ni in 15--25 \msol\ SN explosion is 0.05--0.13~\msol\ 
\citep{woo95, thi96}. Therefore, the Fe ejecta detected in 
\fe164 emission is less than one percent of the total Fe ejecta. 
On the other hand, the observed 
Fe mass corresponds to H (+He) mass of 0.27 $M_\odot$ for 
the solar abundance, which implies that 
the mass of the shocked CSM comprising the Fe filaments should be 
a tiny fraction of the swept-up CSM too.

%Photoionized region in principle 
%can show such high ratios too if it has 
%a thick region of partially ionzed gas. The \fetwo emission 
%in Crab is known to be from such photoionized filaments \cite{gra90}.
%In \g11p2, the spatial coincidence of 
%\fetwo-SE and -NW filaments with the SNR radio shell 
%gives little doubt that they are from the shocked gas.  

%For the above mass loss rate, the H-nuclei density of the smooth wind 
%at the position of \htwo filaments ($r\approx 3$~pc) would be  
%$n_{w}={ {\dot M} / 4 \pi r^2 v_w m_{\rm H}} \approx 0.7$ cm$^{-3}$, which 
%is consistent with the estimation based on the $C$-shock requirement above.
%If the distance to the ambient shock is indeed 7 pc, then 
%essentially all of the wind mass should have been swept up.
%The current shock speed 
%Since the current expansion velocity of the SNR shell is 
%$\vexp=1,350\pm280$~\kms\ according to radio (20 cm) expansion studies 
%by \cite{tam03}, 
%this implies a density contrast  
%between the clump and the smooth wind of 
%$\approx (\vexp/\vc)^2$=1,000--2,000 where $\vc=$30--40~\kms\ is 
%the shock speed in the clump. 
%This implies a H-nuclei 
%density of $\simlt 0.1$~cm$^{-3}$ for the smooth wind.

\section{Conclusion}

G11.2$-$0.3 has been known as an evolved version of Cas A, both 
being SN IIL/b with a significant mass loss before explosion.
Our \htwo results confirm that \g11p2 is 
indeed interacting with a clumpy circumstellar wind as in Cas A.
Clumps with a density contrast of $\sim 3,000$ may be 
common in presupernova circumstellar wind of SN IIL/b.
As far as we are aware, \g11p2 is the first source where 
the presupernova wind clumps are observed in \htwo emission.
The \htwo filament in the northeast 
is of particular interest because it could 
provide a strong evidence for an ambient shock beyond the bright
radio shell. Future spectroscopic studies will reveal the nature of this
filament.  

The \fetwo filaments in \g11p2 are probably composed 
of both shocked CSM and shocked ejecta. The one in the southeast is the
brightest among the known \fe164 filaments associated with  
SNRs and is thought to be 
where the ejecta is heavily interacting with dense CSM.
We note that RCW 103, which is another young 
remnant of SN IIL/b \citep{che05}, 
has a very bright \fetwo filament too. 
The source is similar to \g11p2 in the sense that 
\htwo emission is detected beyond the apparent SNR boundary, although 
the \htwo emission in RCW 103 extends along the entire bright SNR shell 
\citep{oli90}. It is possible that the \fetwo filaments 
in the two remnants are of the same origin. 
The other faint \fetwoo-emitting features of \g11p2 
are thought to be mostly SN ejecta. 
%\cite{ger01} detected 
%\fe164 lines have been detected from ejecta in Cas A and 
%Crab too, but the lines are faint in Cas A \citep{ger01} 
%the lines are UV-excited in contrast to \g11p2 \citep{gra90, hes90}. 
The distribution of \fetwo filaments suggests that 
the explosion produced \g11p2 was asymmetric as in Cas A. 
In Cas A, however, Fe ejecta have been observed mainly in X-rays
although faint \fe164 lines have been detected toward 
several fast-moving ejecta knots from spectroscopic observations by 
\cite{ger01}. 
Future detailed spectroscopic studies will help us to understand 
the nature of the \fetwo filaments and knots in \g11p2 
as well as the SN explosion itself.  

\acknowledgements 
We thank Dave Green for providing his VLA images of \g11p2. 
We also wish to thank Chris McKee and Roger Chevalier for their 
helpful comments. 
D-SM acknowledges a Millikan fellowship from California Institute of Technology. 
This work was supported by the Korea Science
and Engineering Foundation (ABRL 3345-20031017).

{}

\clearpage

\begin{deluxetable}{lccccl}
\tabletypesize{\scriptsize}
\tablecaption{Summary of WIRC Imaging Observations of G11.2-0.3
\label{tbl-1_02}}
\tablewidth{0pt}
\tablecolumns{6} \tablehead{
\colhead {} &\colhead{$\lambda_{\rm center}^a$} & 
\colhead {$\Delta \lambda_{\rm equiv}^b$} & \colhead {Exposure} & 
\colhead {} & \colhead {} \\
\colhead {Filter} &\colhead{($\mu$m)} & 
\colhead {($\mu$m)} & \colhead {(s)} & \colhead {NDI$^c$} & \colhead {Date} 
}

\startdata
[Fe II]$^d$ & 1.644 & 0.0252 & 60 & 12 & 2003. 06. 17, 2005. 08. 27 \\
H$_2$ & 2.120 & 0.0329 & 20 & 36 & 2005. 08. 14 \\
Br$\gamma$ & 2.166 & 0.0327 & 30 & 20 & 2005. 08. 28 \\
$K_s$ & 2.150 & 0.312 & 15 & 90 & 2003. 06. 16 \\
$H$-cont & 1.570 & 0.0236 & 30 & 10 & 2005. 08. 14 \\
$K$-cont & 2.270 & 0.0330 & 60 & 12 & 2005. 08. 27 \\

\enddata
\tablenotetext{a} {Wavelength centers from 
http://www.astro.caltech.edu/palomar/200inch/wirc/wirc\_spec.html}
\tablenotetext{b} { Equivalent width, e.g, $\Delta \lambda_{\rm equiv} = 
\int S(\lambda)d\lambda$ where $S(\lambda)$ is the normalized filter response.}
\tablenotetext{c} {NDI represents ``number of dithered images."}
\tablenotetext{d} {
There are two observations with [Fe II] filter: one in 2003 and the other in 2005.
The observing parameters are the same for the two observations.}

\end{deluxetable}

\begin{deluxetable}{lllll}
\tabletypesize{\scriptsize}
\tablecaption{Detected Lines and Their Strengths \label{tbl-1_01}}
\tablewidth{0pt}
\tablecolumns{5} \tablehead{
\colhead{Position} & \colhead {Wavelength$^a$} &\colhead {Transition} & 
\multicolumn{2}{c}{\underbar{Relative Strength}$^b$} \\ 
\colhead {} & \colhead {} & \colhead {} & \colhead {Observed} & \colhead {Dereddened}
}

\startdata
\fetwo-pk1$^c$ & 1.2567 & \fetwo a $^4D_{7/2}\rightarrow a ^6D_{9/2}$ & 0.314 (0.010)  &  1.04 \\ 
&1.5335 & \fetwo a $^4D_{5/2}\rightarrow a ^4F_{9/2}$ & 0.116 (0.004)  &  0.151 (0.005) \\ 
&1.5995 & \fetwo a $^4D_{3/2}\rightarrow a ^4F_{7/2}$ & 0.102 (0.003)  &  0.113 (0.003) \\ 
&1.6436 & \fetwo a $^4D_{7/2}\rightarrow a ^4F_{9/2}$ & 1.0   &  1.0   \\ 
&1.6638 & \fetwo a $^4D_{1/2}\rightarrow a ^4F_{5/2}$ & 0.052 (0.002) &  0.050 (0.002) \\
&2.1661  & H 4-7 Br$\gamma$                           & 0.030 (0.004) &  0.013 (0.002)  \\
\htwoo-pk1$^d$ & 2.0735 & \htwo       (2-1) S(3)                   & 0.13 (0.01) &  0.14 (0.01) \\
&2.1218 & \htwo       (1-0) S(1)                      & 1.0   &  1.0   \\

\enddata
\tablenotetext{a} {Rest wavelengths of the identified lines.}
%[Fe II] line wavelengths: Quinet et al. 1996
%Br-G: Allen
%H2: CLOUDY
\tablenotetext{b} {{Line} fluxes relative to the \fe164 flux for 
\fetwoo-pk1 and relative to \h212 flux for \htwoo-pk1. The numbers in parentheses are 
$1\sigma$ statistical errors. 
The observed \fe164 surface brightness at \fetwoo-pk1 is 
$1.9(0.2) \times 10^{-3}$ erg cm$^{-2}$ s$^{-1}$ sr$^{-1}$ and 
\h212 surface brightness at \htwoo-pk1 is 
$3.0(0.3)\times 10^{-4}$ erg cm$^{-2}$ s$^{-1}$ sr$^{-1}$ 
according to our narrow-band imaging photometry.}
\tablenotetext{c} {The coordinate of the \fe164 peak position is 
(18$^{\rm h}$ 11$^{\rm m}$ 34.$^{\rm s}$76, $-19^\circ$ 26$'$ 30.$''$0). 
The slit was slightly off from the peak position and 
the spectrum was extracted from a $3''\times 1''$ area (P.A.=38$^\circ$) centered at 
$(\Delta \alpha,\Delta \delta)=(+1.''4\pm 0.''2,-0.''8\pm 0.''2)$ from the peak position
(see Fig. 3).} 
\tablenotetext{d} {The coordinate of the \h212 peak position is 
(18$^{\rm h}$ 11$^{\rm m}$ 32.$^{\rm s}$26, $-19^\circ$ 27$'$ 10.$''$5). 
The slit was slightly off from the peak position and 
the spectrum was extracted from a $3''\times 1''$ area (P.A.=59$^\circ$) centered at 
$(\Delta \alpha,\Delta \delta)=(+0.''5\pm 0.''5,-0.''5\pm 0.''5)$ from the peak position
(see Fig. 9).} 

\end{deluxetable}

\clearpage

\begin{figure}
\epsscale{1.0}
\plotone{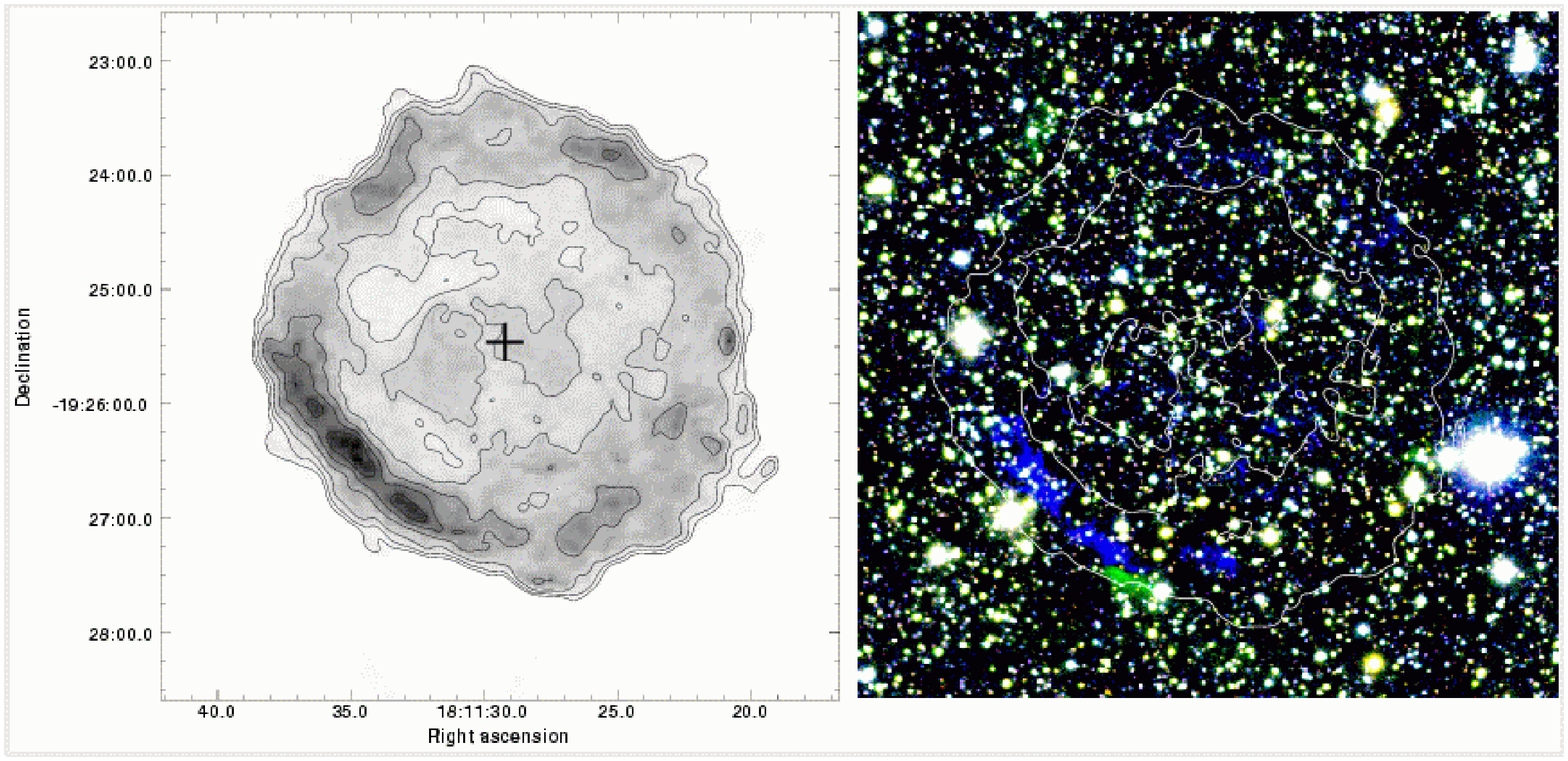}
\caption{Near infrared image compared with radio map of 
\g11p2. Left panel: VLA 1.4 GHz map from \cite{gre88}. The cross 
marks the position of pulsar. 
Contour levels are 0.5, 1, 2, 3, 4, 6, and 8 mJy/pixel where the pixel size 
is $1.''4\times 1.''4$.
Right panel: Three color image generated from \fe164 (B), \h212 (G), and
\brg 2.166 $\mu$m (R). The 
2 mJy/pixel radio contours are overlaid to mark the SNR shell.}
\end{figure}
\clearpage

\begin{figure}
\epsscale{0.70}
\plotone{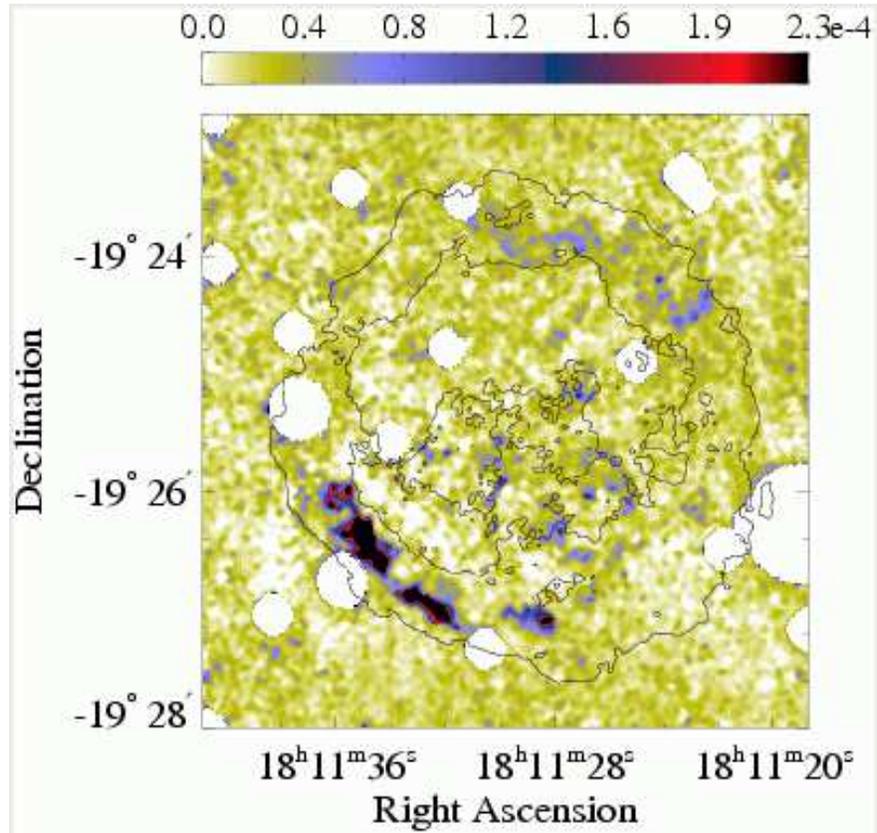}
\caption{\fe164 image with stars subtracted 
using PSF photometry.
Bright stars are masked out and a median smoothing filter is applied to 
remove faint stars (see text for details).
Color bar is given 
at the top of the image 
in units of erg cm$^{-2}$ s$^{-1}$ sr$^{-1}$. The 
2 mJy/pixel radio contours are overlaid. }
\end{figure}
\clearpage

\begin{figure}
\epsscale{0.7}
\plotone{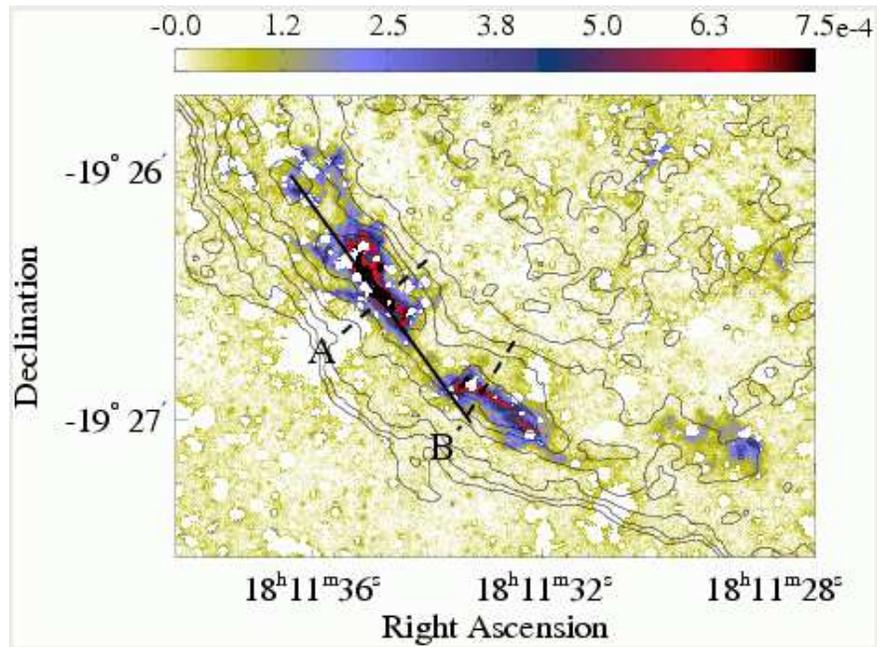}
\caption{An enlarged \fe164 image of the \fetwoo-SE filament. 
Stars are just masked out using a continuum image in order 
to avoid any artifacts in the PSF star-subtraction. Color bar 
is given at the top of the image 
in units of erg cm$^{-2}$ s$^{-1}$ sr$^{-1}$. The thick solid line
shows the slit position for spectroscopy, while the thick dashed 
lines labeled `A' and `B' mark the cuts over which the intensity 
profiles in Fig. 7 are obtained.  
Radio contours are overlaid. }
\end{figure}
\clearpage

\begin{figure}
\epsscale{0.7}
\plotone{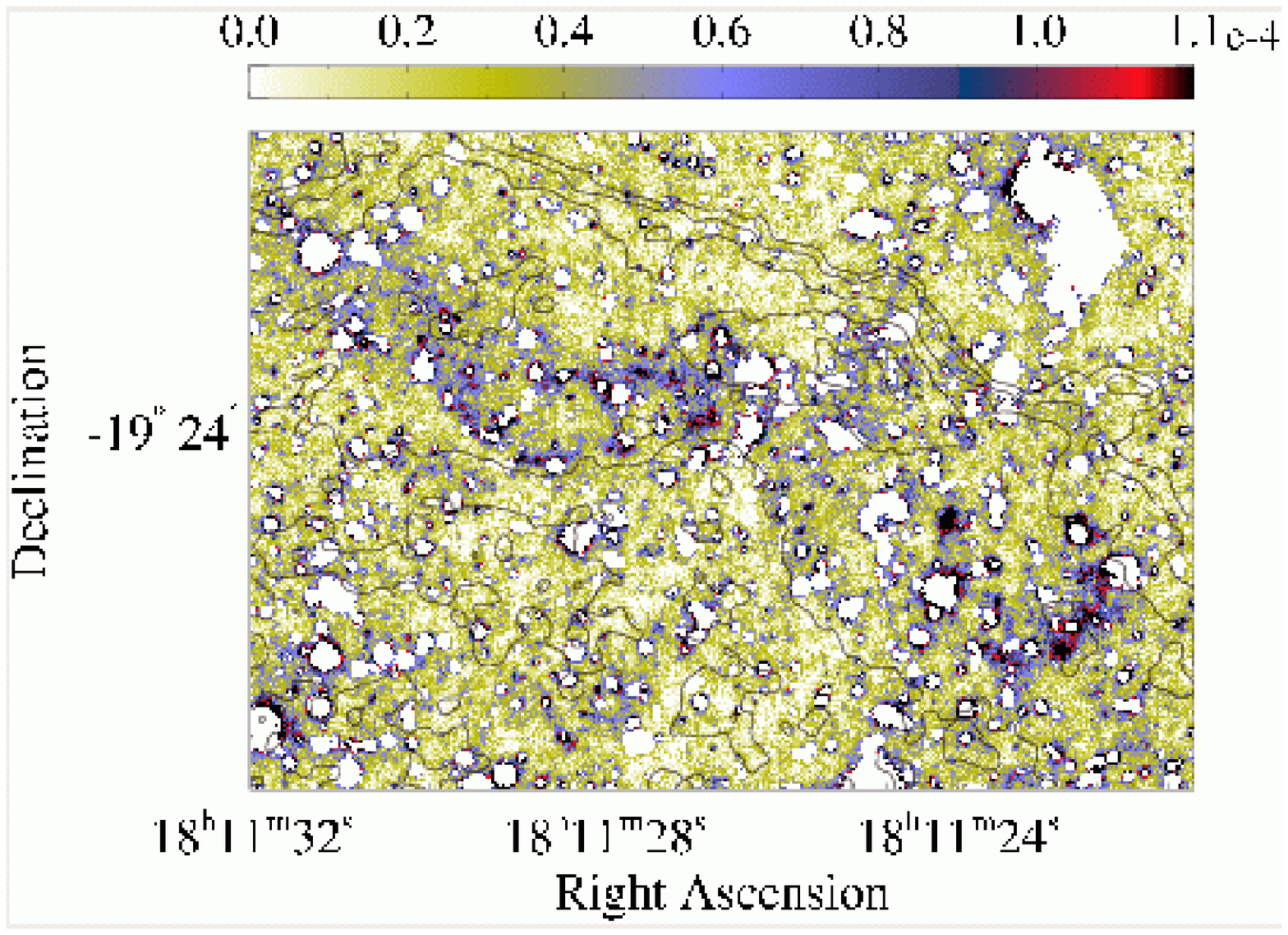}
\caption{An enlarged \fe164 image of the \fetwoo-NW filament. See
the caption in Fig. 3 for explanation.}
\end{figure}
\clearpage

\begin{figure}
\epsscale{0.7}
\plotone{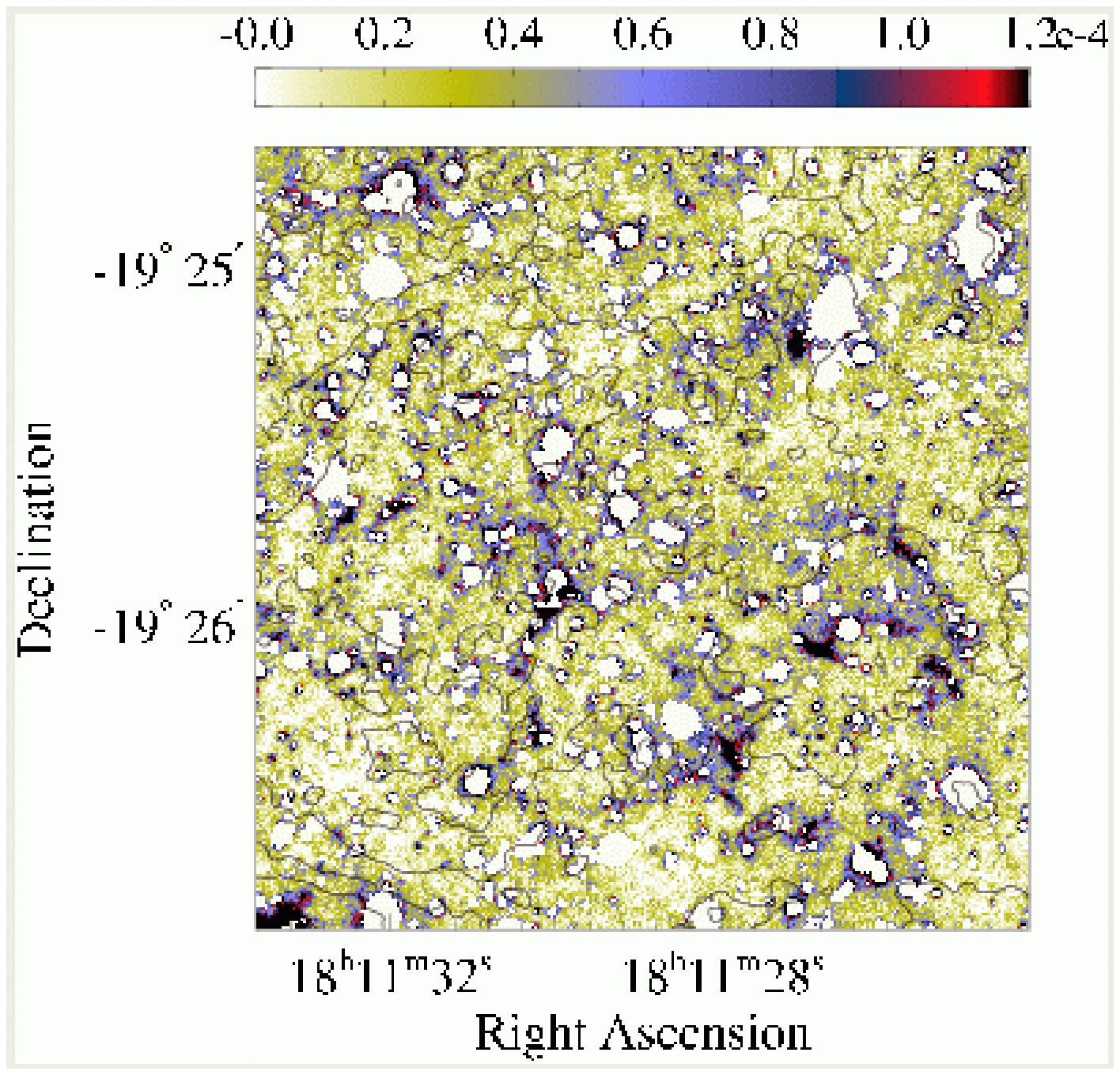}
\caption{An enlarged \fe164 image of the central SNR area. See
the caption in Fig. 3 for explanation.}
\end{figure}
\clearpage

\begin{figure}
\epsscale{0.7}
\plotone{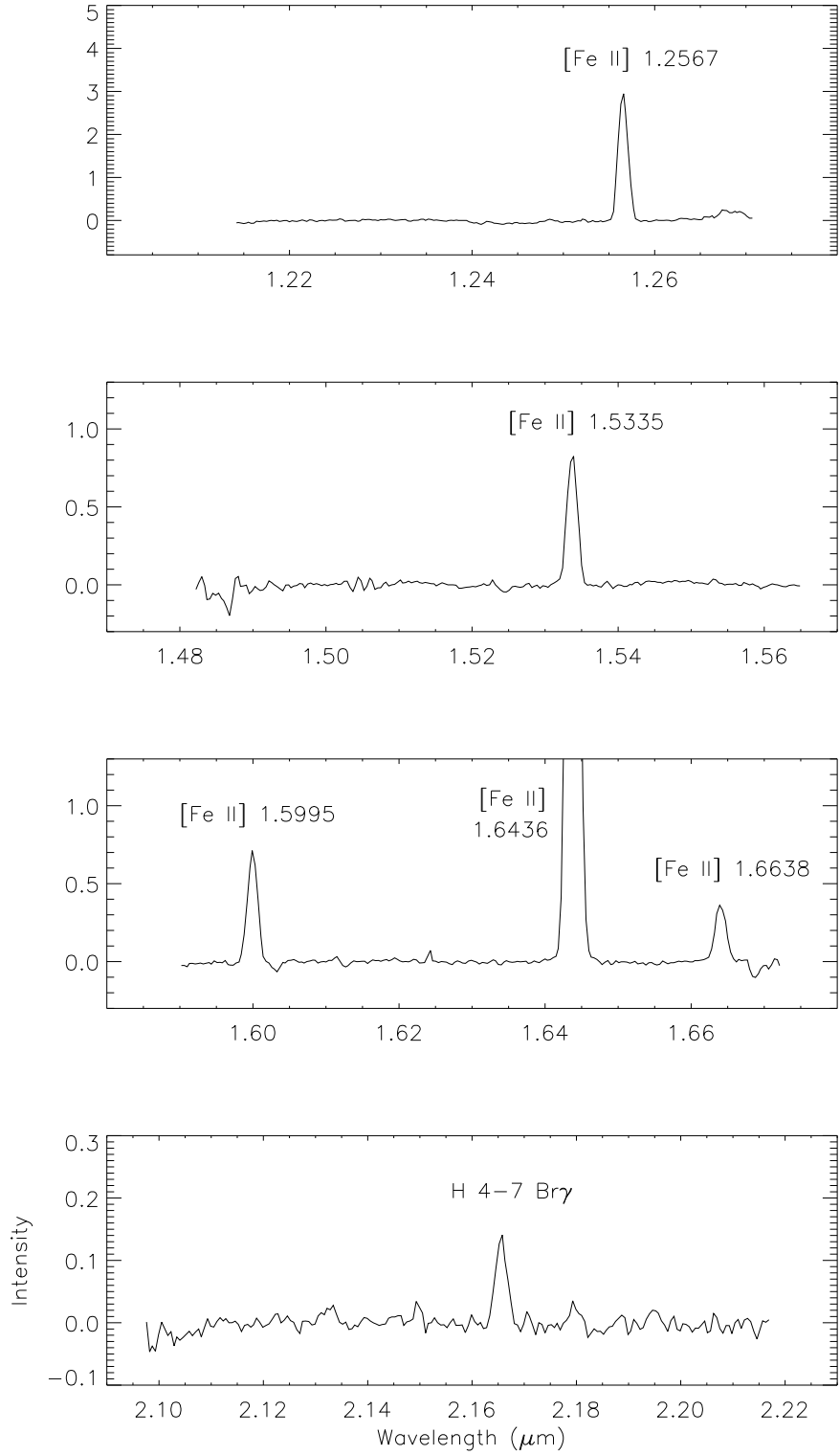}
\caption{\fetwo and \brg spectra of \fetwoo-pk1.
The intensity is in arbitrary scale. The peak 
intensity of the \fe164 line is 6.9 in this scale.}
\end{figure}
\clearpage

\begin{figure}
\epsscale{1.0}
\plotone{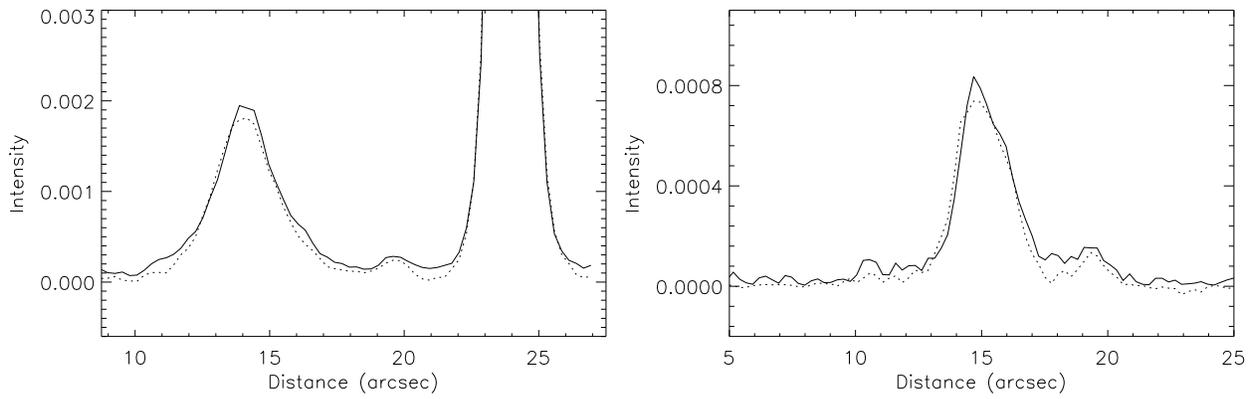}
\caption{Comparison of the brightness distribution of 
the \fetwoo-SE filament in 2005 (solid lines) and that in 2003 (dotted lines). 
Left and right figures show the distribution across the two bright
segments of the \fetwoo-SE filament along the cuts labeled 
`A' and `B' in Fig. 3, respectively. 
The distance in the abscissa is measured from the upper right end of the cuts, 
so that it increases outward from the remnant center. 
Note that the
profiles of the filament in 2005 are
slightly shifted outward from those in 2003. 
}
\end{figure}
\clearpage

\begin{figure}
\epsscale{0.8}
\plotone{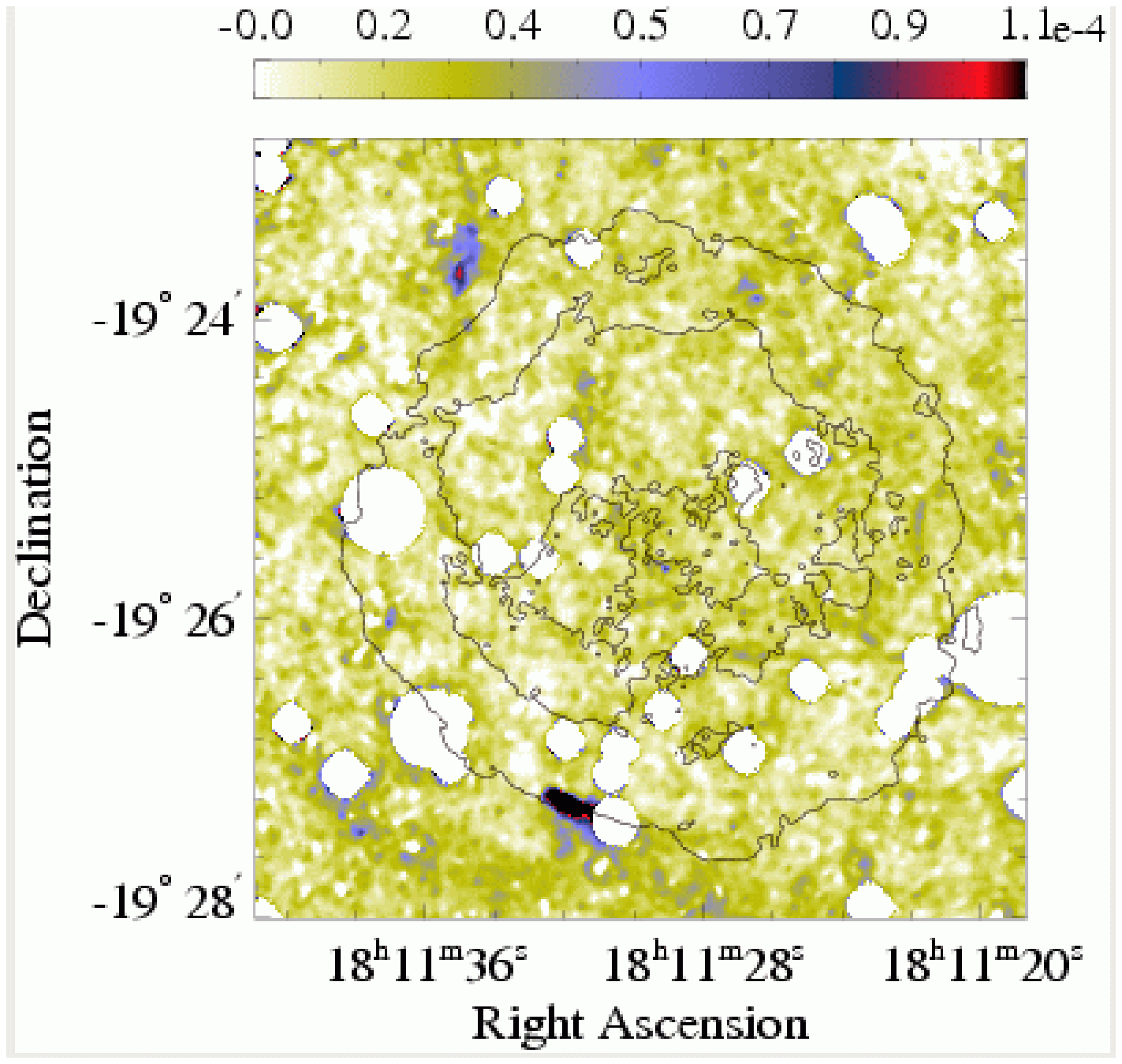}
\caption{Same as Fig. 2, but for \htwo emission.
}
\end{figure}
\clearpage

\begin{figure}
\epsscale{0.7}
\plotone{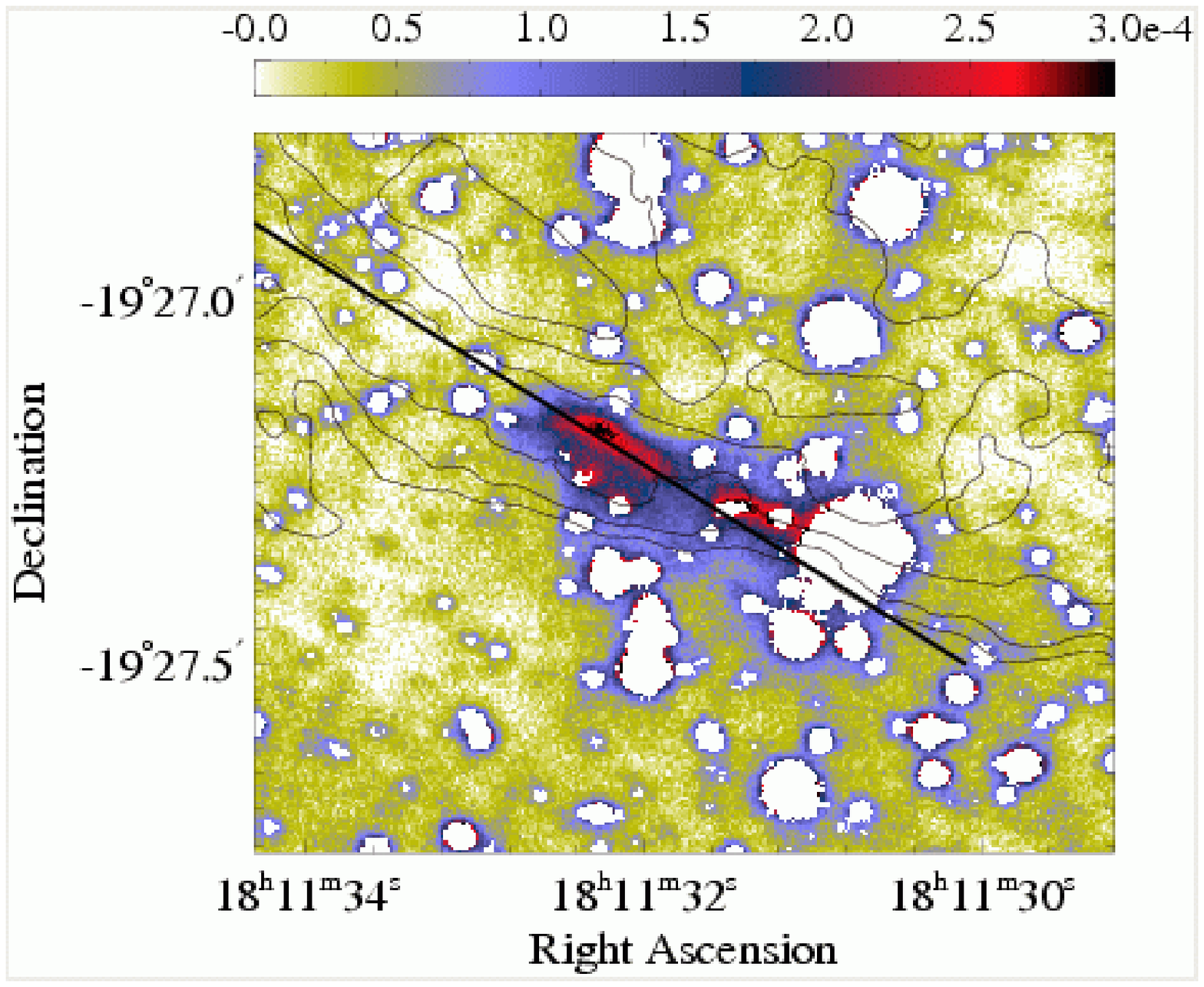}
\caption{An enlarged \h212 image of the \htwoo-SE filament. 
See the caption in Fig. 3 for explanation.}
\end{figure}
\clearpage

\begin{figure}
\epsscale{0.7}
\plotone{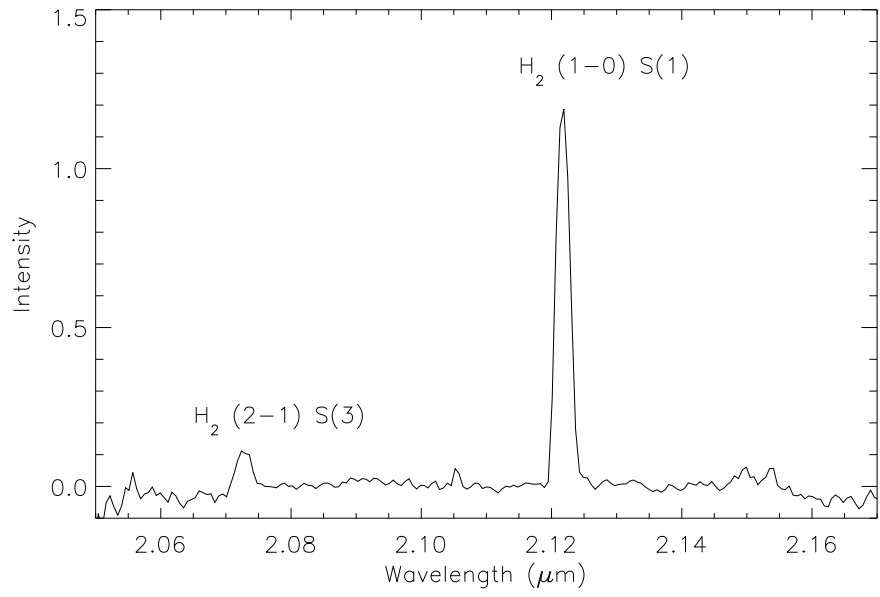}
\caption{\htwo spectra of \htwoo-pk1.
The intensity is in arbitrary scale.}
\end{figure}
\clearpage

\end{document}